\documentclass[12pt]{iopart}

\usepackage{iopams}
\usepackage{graphicx}
\usepackage{multicol}

\newcommand{\unt}[1]{\ensuremath{\,\rm{#1}}}
\newcommand{\degree}[1]{\ensuremath{{#1}^{\circ}}}

\begin{document}

\rapid[Atom Chip for BEC Interferometry]{Atom Chip for BEC Interferometry}

\author{R J Sewell\(^1\)\footnote{Present address:  ICFO-Institut de Ciencies Fotoniques, Mediterranean Technology Park, 08860 Castelldefels (Barcelona), Spain.}, J Dingjan\(^1\), F Baumg\"artner\(^1\), I Llorente-Garc\'{i}a\(^1\), S Eriksson\(^1\)\footnote{Present address: Department of Physics, Swansea University, Singleton Park, Swansea SA2 8PP, United Kingdom.}, E A Hinds\(^1\),G Lewis\(^2\), P Srinivasan\(^2\), Z Moktadir\(^2\), C O Gollasch\(^2\), and M Kraft\(^2\)}
\address{\(^1\)Centre for Cold Matter, Blackett Laboratory, Imperial College,
Prince Consort Road, London SW7 2BW, United Kingdom}
\address{\(^2\)School of Electronics and Computer Science, University of Southampton,
Highfield, Southampton, SO17 1BJ,United Kingdom}
\ead{ed.hinds@imperial.ac.uk}

\begin{abstract}
We have fabricated and tested an atom chip that operates as a matter wave interferometer. In this communication we describe the fabrication of the chip by ion-beam milling of gold evaporated onto a silicon substrate. We present data on the quality of the wires, on the current density that can be reached in the wires and on the smoothness of the magnetic traps that are formed. We demonstrate the operation of the interferometer, showing that we can coherently split and recombine a Bose–Einstein condensate with good phase stability.
\end{abstract}

\submitto{Journal of Physics B: Atomic, Molecular and Optical Physics}

\maketitle
% \tableofcontents

Atom chips are microfabricated devices that control electric, magnetic and optical fields in order to trap and manipulate cold atom clouds and Bose-Einstein condensates (BECs) \cite{reichel1999,hinds1999,folman2002,fortagh2007}. Such devices have significant potential for applications in sensing, metrology and quantum information processing. Although BEC was first created on an atom chip 10 years ago \cite{hansel2000}, the fabrication of functional devices has posed significant technical challenges that have only recently been overcome. In particular, BEC interferometry on an atom chip has been demonstrated using static magnetic fields in combination with radio-frequency \cite{schumm2005}, optical \cite{wang2005} or microwave fields \cite{bolh2009}.

In this paper we report on the fabrication and initial testing of a working BEC interferometer. We have fabricated test batches of atom chips using a variety of techniques for depositing gold on silicon and for etching the required wire structures. After evaluating the quality of these we settled on a fabrication method using electron beam evaporation of a gold layer followed by ion-beam milling to define the wires. We have used one of the atom chips fabricated in this way to make a BEC interferometer.

\begin{figure}[htbp]
	\centering
	\includegraphics[width=8cm]{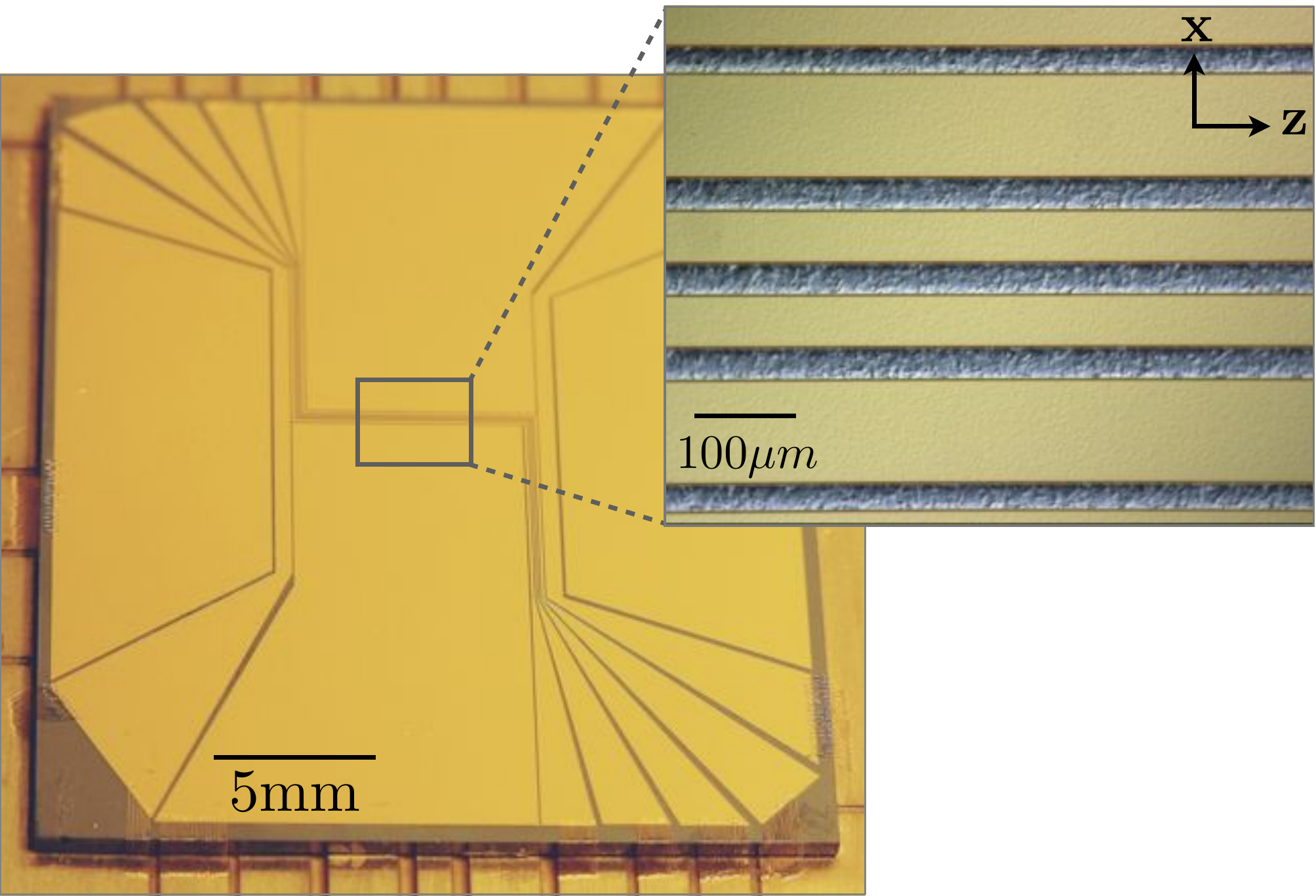}
	\caption{\label{fig:atom_chip}The atom chip used in our interference experiments. Four parallel Z-wires occupy the central region of the chip and there are two additional end wires. The surrounding gold pads form a mirror surface used in pre-cooling the atoms in a MOT. Inset: an optical microscope image of the centre of the chip showing the four parallel trapping wires (in gold). The silicon substrate can be seen in the gaps between the wires (in grey). The roughness of the silicon substrate is due to over-etching during the ion-beam milling.}
\end{figure}

The atom chip that we have fabricated is shown in figure \ref{fig:atom_chip}. Four parallel Z-shaped wires produce the necessary dc and rf fields for trapping and manipulating BECs near the surface of the chip. The wires in the outer pair are \( 100\unt{\mu m} \) wide and have a separation of \( 300\unt{\mu m} \) (centre-to-centre). The inner wires are \( 50\unt{\mu m} \) wide with \( 85\unt{\mu m} \) separation. The central section of the wires along the z-axis, above which the BEC is produced, is \( 7\unt{mm} \) long. Two more wires are patterned onto the chip parallel to the ends of the Z-wires along the x-axis. These are used to provide additional axial trap depth and to adjust the field strength at the trap minimum.

In order to load this chip, cold \( ^{87}\mathrm{Rb} \) atoms from a low-velocity intense source (LVIS) are first captured \( 4\unt{mm} \) from the surface in a magneto-optical trap (MOT) \cite{sinclair2005}. The gold surface serves as a mirror that reflects some of the laser cooling light, allowing the MOT to be formed close to the chip. The atoms are then passed to the magnetic trap, where the cloud is further cooled by forced evaporation using an rf field to eject the most energetic atoms. A BEC is formed at approximately \( 500\unt{nK} \) and this provides the coherent matter wave for our interferometer. The procedure is similar to that described in our previous publications \cite{jones2003}.

Although atom chips are made using standard microfabrication techniques, the experimental requirements impose a number of unusual design constraints. In order to create magnetic traps with sufficient depth, the wires must be able to carry several amperes for a period of ten to twenty seconds. At the same time they should be as small as possible to manipulate the atoms on a small length scale. This means that they must be able to carry very large current densities. The wires must also be very smooth on length scales of up to several hundred microns in order to minimize transverse currents that lead to roughness of the magnetic trap and fragmentation of cold clouds (see below). Finally, in order to facilitate pre-cooling of the atoms in a MOT, the entire chip surface must have a mirror finish.

Gold is our material of choice since it has a low resistivity and high reflectivity at the relevant laser wavelength. We use a silicon substrate in order to take advantage of mature fabrication techniques that will allow for future integration of optical elements such as fibres, cavities, waveguides and pyramid MOTs into a single device \cite{eriksson2005,trupke2006,trupke2007,pollock2009}. Silicon has good thermal conductivity, but must be covered in a thin oxide layer to ensure that the wires are electrically isolated from the substrate.

The wires can be fabricated using standard lithographic techniques or by adapting thin film hybrid technology \cite{reichel2002,fortagh2007}. High quality wires have been made by applying a lift-off technique to an evaporated gold layer \cite{groth2004,kruger2007,trinker2008}. Evaporated gold has a good surface finish, and the lift-off technique gives the wires good edge definition when they are thin. However, the method is not well developed for films as thick as ours. Other groups have patterned a thin gold film photolithographically, then grown thick wires by electrochemical deposition \cite{ott2001,lev2003,esteve2004}. Thicker wires can be fabricated using this technique, but the homogeneity of the gold and definition of the edges are not as good. In our previous work we have studied techniques for fabricating gold wires using sputtered and electrochemically deposited gold layers that are subsequently patterned by wet-etching or ion-beam milling \cite{koukharenko2004}. We have also investigated growing wires by electrochemical deposition into a mould formed by a lithographically patterned photoresist \cite{lewis2005}. In the light of all these studies, we have fabricated our BEC interferometer chip using a thick evaporated gold layer patterned by ion-beam milling.

\begin{figure}[htbp]
	\centering
	\includegraphics[width=8cm]{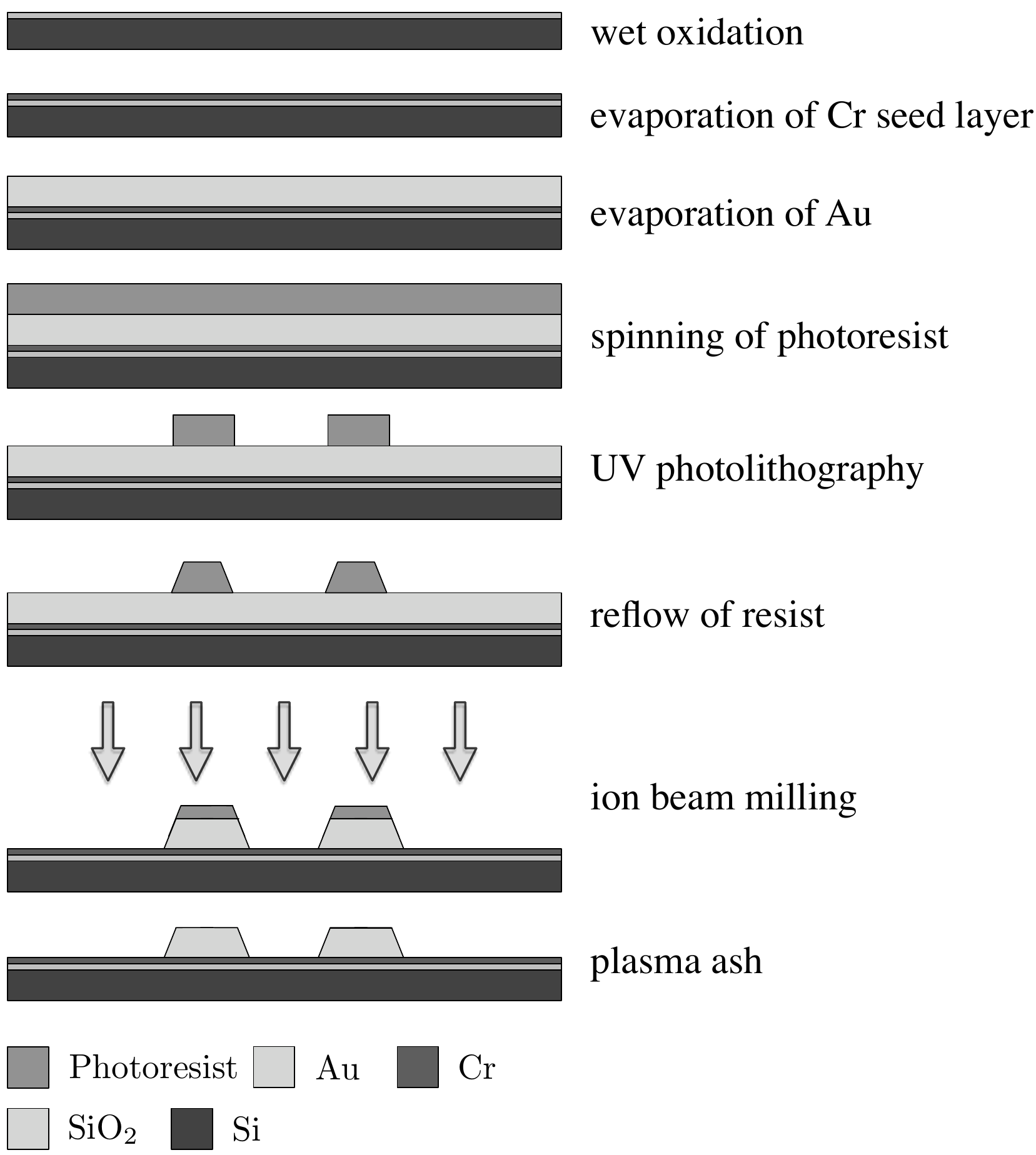}
	\caption{\label{fig:ibmFlowDiagram}Fabrication process flow for electron beam evaporation of a thick gold film followed by ion-beam milling.}
\end{figure}

We use a 4 inch p-doped [100] silicon wafer with a resistivity of \( 17\mbox{-}30\unt{\Omega cm} \) to produce 16 atom chips. The wafer is cleaned using the standard Radio Corporation of America procedures and fuming nitric acid. It then undergoes wet oxidation to produce a \( 100\unt{nm} \) thick oxide layer, the first of several steps illustrated in figure \ref{fig:ibmFlowDiagram}. A \( 40\unt{nm} \) adhesion layer of chromium is then evaporated over the whole surface, followed by \( 3\unt{\mu m} \) of gold deposited in five steps of \( 600\unt{nm} \) to avoid overheating the evaporator. After cleaning in fuming nitric acid, a \( 2.2\unt{\mu m} \) thick layer of HPR504 photoresist is spun onto the gold at \( 500\unt{rpm} \) for 10 seconds, followed by 30 seconds at \( 1500\unt{rpm} \). This is given a soft bake at \( \degree{90}\mathrm{C} \) for 120 seconds, then it is patterned by UV lithography using a Karl Suss MA8 machine for 9 seconds at \( 6.5\unt{mW}\unt{cm^{-2}} \). Finally, a hard bake is done for 30 min at \( \degree{140}\mathrm{C} \) so that the resist will be easier to remove after it has been subjected to ion-beam milling. This also causes the resist to develop sloping sides as it reflows a little.

Milling is done on an IONFAB 300+, with \( 388\unt{V} \) of beam voltage, \( 200\unt{mA} \) of current and \( 276\unt{V} \) of accelerating voltage. The wafer is cooled to a temperature of \( \degree{21}\mathrm{C} \) using helium and is milled for 50 minutes, resulting in a maximum cutting depth of \( 4.4\unt{\mu m} \). The resist is quite hard to remove after exposure to the ions, despite the hard bake, so we use a plasma asher for this purpose run at \( \degree{110}\mathrm{C} \) with \( 600\unt{W} \) for 60 minutes. Once all the resist has been removed the wafer is cleaned in fuming nitric acid. The etch rate is not uniform across the wafer, resulting in over-etching in some places. Where the etch is too deep, the mill can go through the oxide layer and into the silicon substrate itself. In that case, re-deposited silicon on the side walls of the cut makes an electrical short to the wafer. This debris is removed by a 5 second buffered HF acid dip (7:1) followed by a 5 minute KOH etch. Finally, we use a diamond scriber to cleave the wafer into 16 separate atom chips \( 24 \unt{mm} \) wide and \( 26 \unt{mm} \) long.

\begin{figure}[htbp]
	\centering
	\includegraphics[width=8cm]{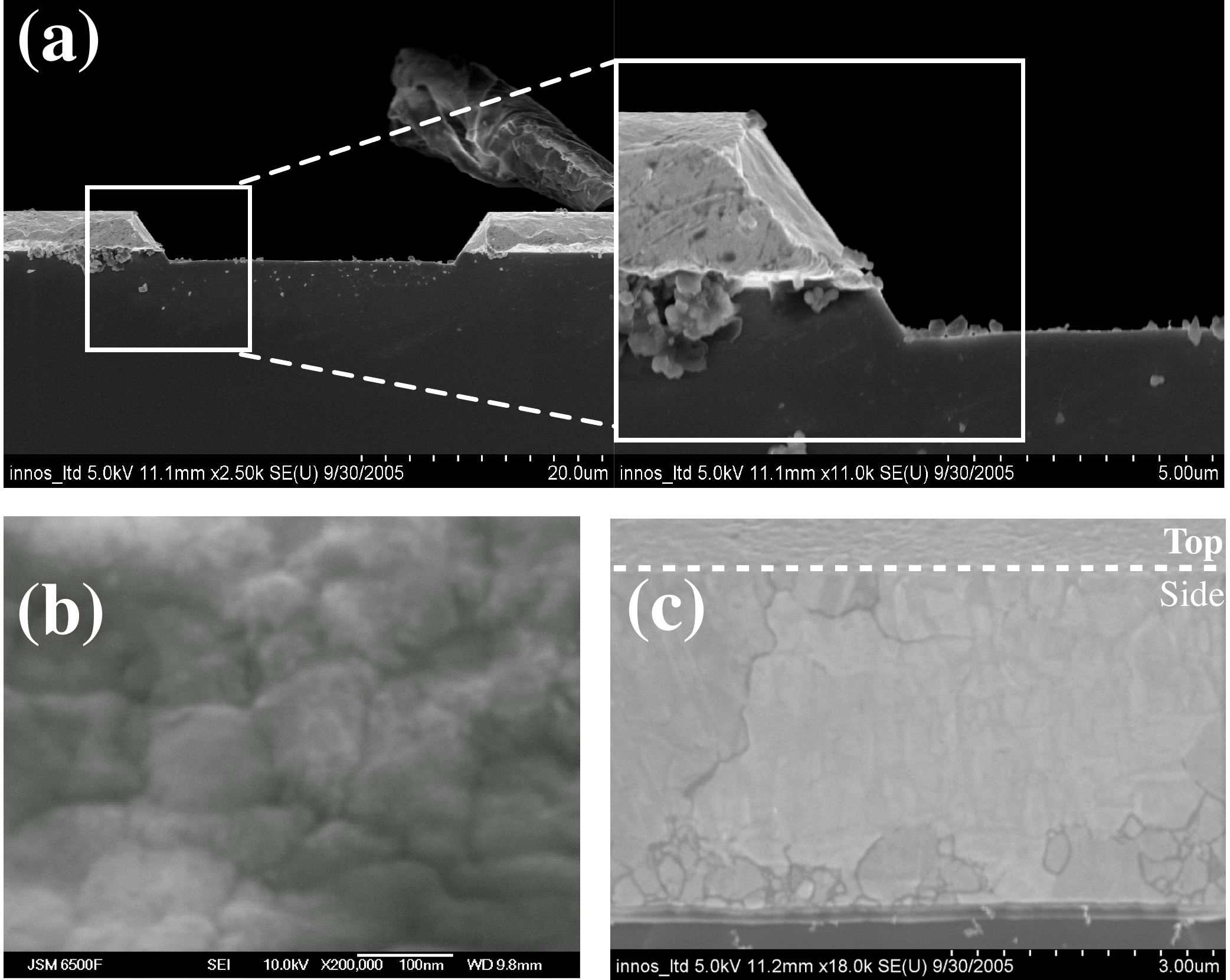}
	\caption{\label{fig:sem}SEM images of the gold wires fabricated by e-beam evaporation followed by ion-beam milling. Image (a) Wire cleaved through the middle to reveal sloping sides. (b) Top surface of the wire. (c) View facing the sloping side wall of the wire.}
\end{figure}

Cleaving a chip through the middle allowed us to examine the cross sectional profiles of the wires using a scanning electron microscope (SEM). Figure \ref{fig:sem}(a) shows the sloping side walls of the wire, transferred from the resist to the wire by erosion of the resist during the milling process. One also sees that the milling was too deep on this wire and penetrated into the silicon. Figure \ref{fig:sem}(b) shows an SEM image of the surface of one of the gold wires, which was found using an atomic force microscope to have \( 3\unt{nm} \) RMS roughness. Figure \ref{fig:sem}(c) shows an SEM image of the sloping side wall of one of the gold wires. Some grain structure is evident on the \( \unt{\mu m} \) scale, but there is no sign of any layering due to the multi-stage evaporation. The surface and wire edge are smooth on the scale of this image.

The maximum usable current density in the chip wires follows from the temperature rise due to resistive power dissipation and is limited by thermal conduction. The insulating \( \rm{SiO}_{2} \) layer is the main barrier to heat flow. When current is turned on, the wire temperature rises rapidly over some microseconds until the drop across this layer saturates. Thereafter, the wire temperature rises more slowly, as determined by thermal conduction into the silicon substrate and on into the mounting structure, made of oxygen-free copper embedded in a Shapal-M (AlN) base plate, connected to an 8 inch stainless steel vacuum flange. The mounting structure acts as a heat sink.

The wires were tested by passing current through them and using the change in resistance to monitor the slow temperature rise. Taking an increase of \( \degree{150}\mathrm{C} \) (50\% increase in resistivity) as a reasonable working upper limit, we measured maximum current densities of \( 8.8\times10^{9}\unt{A}\unt{m^{-2}} \) in the \( 50\unt{\mu m} \) wide wires and \( 6.1\times10^{9}\unt{A}\unt{m^{-2}} \) in the \( 100\unt{\mu m} \) wide wires with current pulses ten to twenty seconds long and with the atom chip in vacuum.

\begin{figure}[htbp]
	\centering
	\includegraphics[width=8cm]{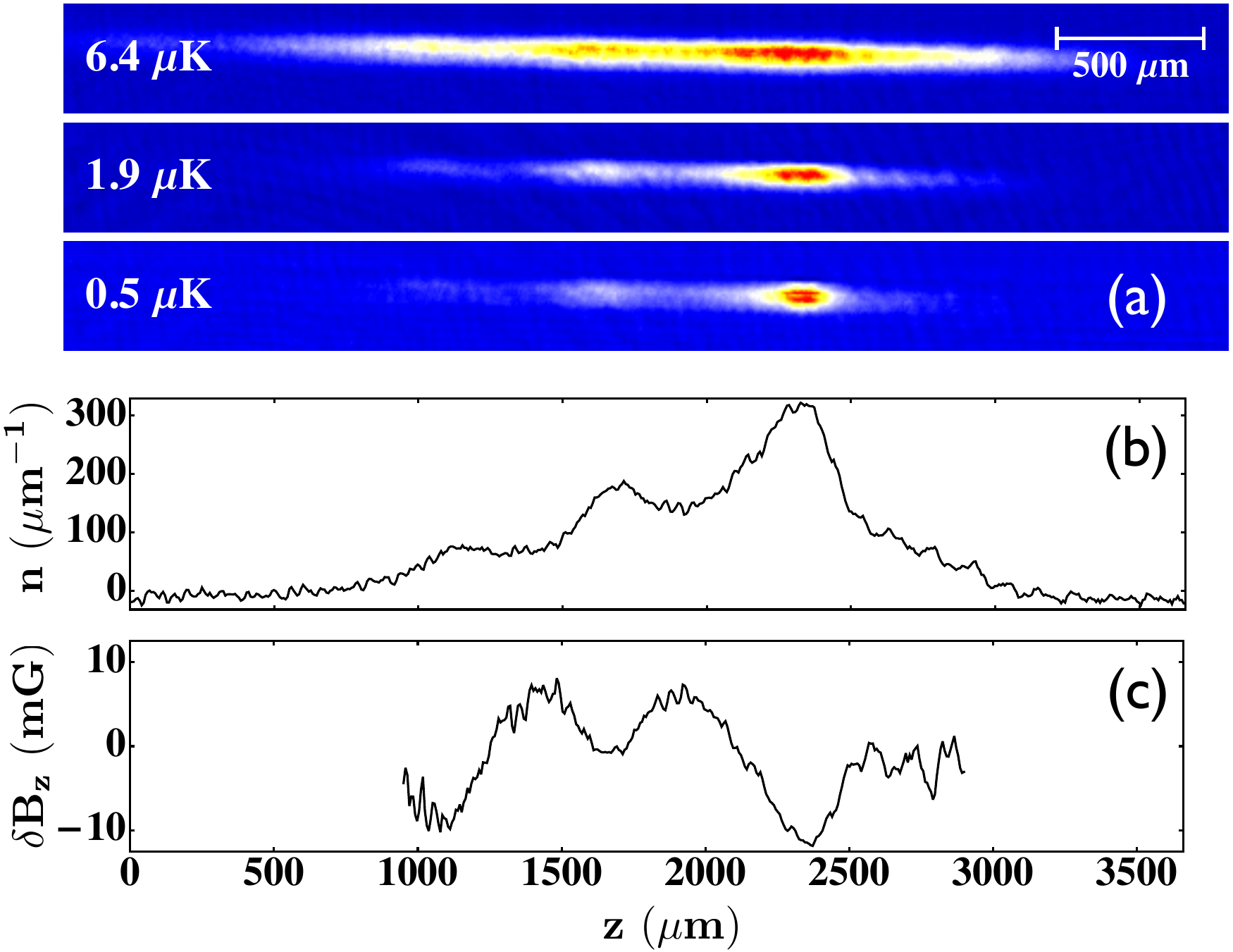}
	\caption{\label{fig:bec}Cold atom studies of the chip. (a) Absorption images of successively colder clouds taken after turning off the trap and accelerating the clouds away from the chip for \( 3\unt{ms} \).  At \( \sim 2\mbox{-}3\unt{\mu K} \) the cloud begins to sense roughness of the trapping potential, and the wing to the left becomes distinct from the main cloud. A BEC begins to form in the largest of these lumps at \( 500\unt{nK} \). (b) Linear number density \( n(z) \) of the \( 1.9\unt{\mu K} \) cloud. (c) Inferred deviation of trap from a smooth harmonic potential (\( \omega_{z}=2\pi\times6.5\unt{Hz} \)), expressed as the magnetic field \( \delta B_{z} \) that causes it.}
\end{figure}

Atoms are loaded into the chip by passing them from the MOT to a magnetic trap at a height \( y \simeq 150\unt{\mu m} \) above one of the wires. This is formed by passing \( 2\unt{A} \) through the wire, with a bias field of \( B_{x} = 24.8\unt{G} \). We then cool the atoms to the BEC transition by forced rf evaporation. The absorption images in figure \ref{fig:bec}(a) show the cloud at several temperatures near the end of the evaporation process. At \( 6\unt{\mu K} \) the cloud is roughly \( 1\unt{mm} \) long and exhibits an extended wing in the left hand side. With further cooling down to \( 1.9\unt{\mu K} \) that wing becomes a clearly separate cloud, due to a subsidiary minimum in the axial potential. As described in \cite{jones2004}, we can derive the variation of the potential from the density distribution of these atoms, shown in figure \ref{fig:bec}(b). The result is illustrated in figure \ref{fig:bec}(c). This roughness is due to transverse currents \( \delta I_{x}(z) \), which generate fields \( \delta B_{z}(z) \) parallel to the wire. Since the bottom of the trapping potential is set by \( B_{z}(z) \), these fields make the trap rough \cite{esteve2004,schumm2005b,kruger2007,jones2004}. The angular variation of the current can be estimated from the ratio of the noise field to the main field, which is approximately \( \pm 10^{-4} \). Since the variation takes place over typically \( \pm 200\unt{\mu m} \), the centre of the wire need only deviate by \( 20\unt{nm} \) over this length to cause the effect that we see. It seems probable that this is due to slight variations in the edges of the wire, though it could also be due to minor defects in the homogeneity of the gold. It would be interesting to see if this could be improved by omitting the reflow of the resist to achieve better definition of the edges. The magnitude of the potential variation at this distance is similar to that seen in electroplated wires of similar dimensions and larger than that reported in evaporated wires patterned using a lift-off technique \cite{kruger2007}.

\begin{figure}[htbp]
	\centering
	\includegraphics[width=8.5cm]{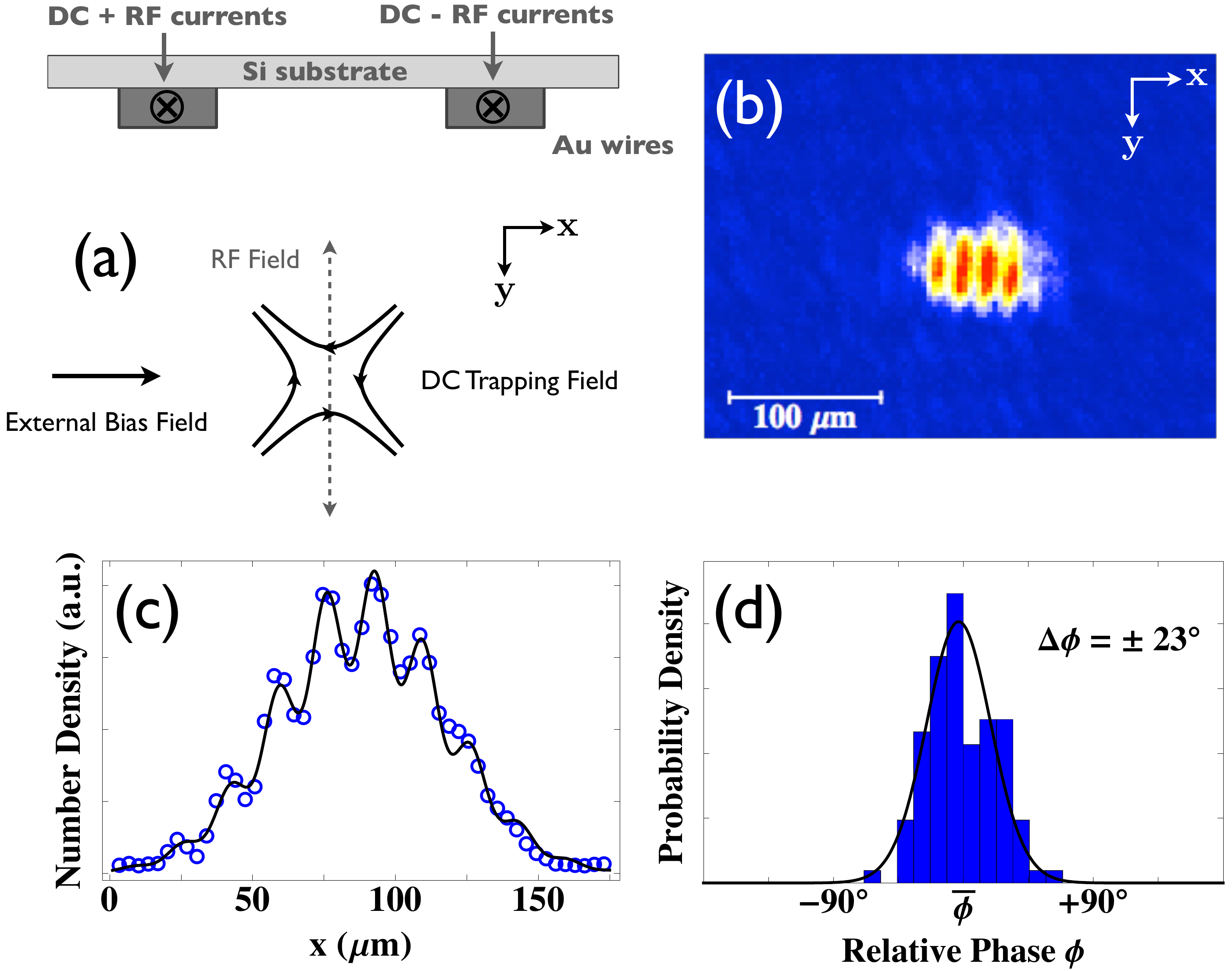}
	\caption{\label{fig:interference}(a) Configuration of static and rf fields used to split the BEC coherently. (b) Absorption image of the atomic cloud showing interference fringes formed when the trapping potential is turned off and the two arms of the BEC interferometer are allowed to overlap in free fall. (c) The relative phase is obtained by fitting a modulated gaussian (solid line) to a slice through the centre of the interference pattern. (d) Histogram of the relative phases extracted from 103 repetitions of the experiment. The solid line is a fit to the data using a normal distribution. The standard deviation is \( \Delta\phi=\pm23^{\circ} \).}

\end{figure}

In order to split the matter wave with our atom chip, we alter the potential by adding near resonant rf fields as proposed by \cite{zobay2001,zobay2004} and demonstrated by \cite{schumm2005,colombe2007,jo2007}. The experimental arrangement is shown in figure \ref{fig:interference}(a). Two wires carrying parallel dc currents form a 2D quadrupole with the help of the bias field. We evaporate to BEC in this trap and continue the evaporation until no discernible thermal atoms remain, at which point the BEC has \( \sim 1.5 \times 10^4 \) atoms and a chemical potential \( \mu = h \times 3\unt{kHz} \). The addition of rf currents, \( \degree{180} \) out of phase, generates a near-resonant rf field along \( y \) that splits the cigar-shaped cloud into two parallel clouds. The trap can be smoothly deformed from a single to a double well by ramping the intensity and/or frequency of the rf. A typical double well used in our interference experiments has a separation of \( \sim4\unt{\mu m} \) between the two trap minima, and a barrier height of \( \sim10\unt{kHz} \).

After allowing the two parts to evolve separately for approximately \( 1\unt{ms} \), we read out the relative phase between them by turning off the trapping potential and allowing them to overlap in free fall. We then take an absorption image of the density distribution, which exhibits interference fringes perpendicular to the splitting axis, as illustrated in figure \ref{fig:interference}(b). We analyse the pattern by fitting a modulated gaussian \( n(x) = g(x)\left(1+\alpha\cos\left(\frac{2 \pi x}{\Lambda} + \phi\right) \right) \) to a slice through the centre, as shown in figure \ref{fig:interference}(c), to determine the relative phase \( \phi \). In figure \ref{fig:interference}(d) we plot a histogram of the phases extracted from 103 repetitions of the experiment. The standard deviation of these is \( \pm23^{\circ} \), indicating that the splitting produces a well-defined initial relative phase between the two arms of the interferometer, as is required for a useful measuring device. This phase spread is similar to that reported by Schumm et al. \cite{schumm2005b} for similar experimental parameters and evolution time.

In conclusion, we have fabricated an atom chip by a process involving electron beam evaporation of a thick gold layer on a silicon substrate followed by ion-beam milling. The resulting wires are able to carry high density dc and rf currents and are sufficiently smooth and uniform to trap a cold atom cloud close to the surface of the chip. We have used one of these atom chips to make a working BEC interferometer with good phase stability.

\section*{Acknowledgements}
\addcontentsline{toc}{section}{Acknowledgements}
The authors acknowledge the expert technical assistance of Jon Dyne. This work was supported by the UK EPSRC, by the Royal Society, and by the European Commission through the SCALA and AtomChips networks.

\section*{References}
\addcontentsline{toc}{section}{References}

\end{document}